\documentclass[11pt]{article} %
\usepackage[utf8]{inputenc}
\usepackage{amsmath}
\usepackage{amsfonts}
\usepackage{float}
\usepackage{graphicx}

\usepackage{enumerate} %
\usepackage{xurl}
\usepackage{setspace}
\usepackage{hyperref} %
\usepackage{caption}
\hypersetup{hidelinks = true} %

\usepackage{csquotes}
\usepackage[notes,isbn=false,maxcitenames=2, backend=biber]{biblatex-chicago}

\bibliography{community}

\usepackage{geometry}
\usepackage{xcolor}
\usepackage{graphicx}

\usepackage{tikz}
\usetikzlibrary{tikzmark}
\usetikzlibrary{arrows.meta}
\usetikzlibrary{calc}

\usepackage{algorithm}
\usepackage{algpseudocode}

\author{Dominik J. Schindler%
\thanks{Department of Mathematics, Imperial College London, UK;  \ 
    \small\texttt{dominik.schindler19{\scriptsize @}imperial.ac.uk}}
\and Matthew Fuller%
\thanks{Department of Media, Communications and Cultural Studies, Goldsmiths, University of London, UK;
\ 
    \small\texttt{m.fuller{\scriptsize @}gold.ac.uk}}   
    }

\title{%
Community as a Vague Operator: Epistemological Questions for a Critical Heuristics of Community Detection Algorithms
}

\begin{document}
\maketitle

\begin{abstract}
\noindent
In this article, we aim to analyse the nature and epistemic consequences of what figures in network science as patterns of nodes and edges called `communities'. Tracing these patterns as multi-faceted and ambivalent, we propose to describe the concept of community as a `vague operator', a variant of Susan Leigh Star's notion of the boundary object, and propose that the ability to construct different modes of description that are both vague in some registers and hyper-precise in others,  is core both to digital politics and the analysis of `communities'.  Engaging with these formations in terms drawn from mathematics and software studies enables a wider mapping of their formation.  Disentangling different lineages in network science then allows us to contextualise the founding account of `community' popularised by Michelle Girvan and Mark Newman in 2002. After studying one particular community detection algorithm, the widely-used `Louvain algorithm', we comment on controversies arising with some of their more ambiguous applications. We argue that `community' can act as a real abstraction with the power to reshape social relations such as producing echo chambers in social networking sites. To rework the epistemological terms of community detection and propose a reconsideration of vague operators, we draw on debates and propositions within the literature of network science to imagine a `critical heuristics' that embraces partiality, epistemic humbleness, reflexivity and artificiality.
\end{abstract}


\section{Introduction}

Network science emerges as a term in the late nineteen-nineties and consists of a series of `content agnostic' ways to analyse structures of various kinds as networks or graphs~\autocite{newmanStructureDynamicsNetworks2006}.  It can be understood as a revival of the much older social network analysis through the influence of physics.~\autocite{freemanGoingWrongWay2008} The kind of things network scientists work on range from the structure of proteins, to relations between social media posts, to chains of influence in academic research.  Tools and approaches from network science are also often drawn into other fields, to show connections amongst entities as diverse as members of the ruling class or of criminal trading networks—as developed for instance in the meticulous work of artist Mark Lombardi~\autocite{hobbsMarkLombardiGlobal2004}—or to construct a  taxonomic characterisation of the intestinal microbiota involved in gout.~\autocite{guoIntestinalMicrobiotaDistinguish2016} Work in the field and in the applications of its tools seems to suggest the possibility of finding shared `hidden laws' amongst often very different kinds of formations.  

By the present day, the working vernacular of network visualisations has become a familiar part of contemporary culture.  For instance, Figure~\ref{fig:citation_graph} and Figure~\ref{fig:Example_Network} below typify such images.  They are composed of two types of entity, edges or connecting lines and vertices or dots where two or more lines meet. But what is meant by these patterns of dots and lines?  In network science, the notion of `community' was coined to grapple with these patterns~\autocite{girvanCommunityStructureSocial2002} and `community detection algorithms' such as the `Louvain algorithm' are used today to discriminate such patterns in large networks with millions of nodes and edges.~\autocite{blondelFastUnfoldingCommunities2008} In particular, community detection algorithms can be interpreted as methods for unsupervised machine learning that are supposed to find patterns in data without a given ground truth.~\autocite{hastieElementsStatisticalLearning2009} To delve into these patterns requires asking questions of their meaning: what do they stand in for, what do they signify, and what do they create? Further, what are the ways in which these arrangements of dots and lines, and the calculations that produce them, have potential cultural and political effects? To address this means recognising these patterns as a visual articulation of mathematical relationships. In order to hold these two aspects together, recognising their mutual inherence and differentiation, their particular and conjoint epistemic dimensions need to be addressed. One of the ways to do this is by understanding the way in which the notion of community provides in itself something of a conceptual vertex between different modes of analysis and understanding.  

Since social media have incorporated the form of the graph, without, oddly enough, giving users actual sight of it, social networks have become part of the everyday furniture of social relations, given for instance in the brute facticity of artifacts like the following to follower ratios on Twitter, the commonplace of `virality'~\autocite{sampsonViralityContagionTheory2012} and the social role of the influencer, a social function that is in some ways predicated upon the operation of graphs.  Such graphs play numerous roles.

We move from a society understood, from some disciplinary or technical perspectives, to be composed of individuals in networks that can be analysed by means of reserved or neutral observation to a society of analysis whose givens are networks in which power operations are implemented.  In this set-up it should be of scant surprise that the word community appears as capable of interpreting many kinds of phenomena at the exact point in time when, if it has not entirely vanished, community, in its hitherto understood senses—in the social—seems often to have been mechanised, and often by the very means that redescribe it in more generalisable terms.  In this condition, it is perhaps rather wince-inducing to rifle through the techniques of network analysis to try, not only to understand them, but to evaluate the conditions in which they might be worked.  Nevertheless, there is something fascinating here, and one of the ways of understanding the way these techniques not only address but compose the present is by delving into them.

In this article, we aim to analyse the nature of what figures in network science as a community, trace the historical lineages of community detection algorithms and examine a specific case study of an algorithm for community detection and the notion of community it addresses. We introduce the notion of the `vague operator', a specific kind of boundary object, to describe the various kinds of interplay between the hyper-precise and the vague that are embodied in the conjuncture of community and community detection algorithms. We then look into the broader standing of  heuristics in relation to algorithmic practices and suggest a  `critical' heuristics attuned to the epistemic politics of `vague operators'.

\section{Community / Detection}

\subsection{Lineages of Community Detection Algorithms}\label{sec:Genealogy}

Mathematical practices are interwoven with their historical and technological gestation, but are rarely reducible to them.  Computation in turn has changed mathematical ideas and modes of calculation in multiple ways.~\footnote{It has for instance introduced pathways to certain kinds of mathematical objects whose development only took off with sufficient capacity of calculation. An example would be the development of a renewed interest in what came to be called fractals, (re)emerging with the PCs of the 1980s.~\autocite{mandelbrotFractalistMemoirScientific2013}} The uptake of graph theory for network science purposes coincides with the increased availability of network datasets during the 1990s development of computer networks and the internet~\autocite{newmanStructureDynamicsNetworks2006}---which in some ways become both its metaphor and locus of veridiction, the space where it became true as something natively artificial. To say this is not to claim that mathematics is simply on the receiving end of history, nor of technical histories. Mathematics, as a means of thinking that has great capacity of abstraction also contains some  possibility of thinking outside of historical constraints, of over-leaping them, and in this way may also act as one of their determinants.

Whilst we can take the above considerations into account, the focus of our paper lies on the mathematical practices that have shaped the central concept of community in network science.  A genealogy of community detection needs to disentangle different lineages that have roots in other techniques (not named after community) and run in parallel across disciplines, mostly the social sciences and statistical physics. We can only approximate these lineages due to the enormous amount of publications involved and so present one narrative only, one that is influenced by discussions with different practitioners in network science. A certain amount of reticence is therefore present in this account as we map an initial development in the social sciences and a subsequent, and initially separate, one developed in statistical physics.

In sociology, social network analysis has a twentieth century history, admirably given by Katja Mayer in a 2009 article that traces its links to search engine technologies.~\autocite{mayerSociometrySearchEngines2009}  Mayer argues that social network analysis or sociometry developed alongside related techniques such as citation analysis, formulated as means for measuring authority and participation in academic publishing, techniques that soon became extended as a measure for centrality, opportunities for `self-realisation', cultural significance and optimisation amongst other factors.  This phenomena is also perceptively described by Bernhard Rieder in his account of the genealogy of PageRank.~\autocite{riederWhatPageRankHistorical2012} Aside from this thread of work, the development of methods for what is today called `community detection'  has a longer tradition under different names such as `network partitioning' or `clustering'.~\autocite{fortunatoCommunityDetectionGraphs2010} One important predecessor from social network analysis is the mathematically simpler concept of a graph `clique',~\autocite{wassermanSocialNetworkAnalysis1994} defined as a set of nodes of which each pair of nodes is connected in the graph. This concept was used by Duncan Luce and Albert Perry in 1949 to algorithmically obtain group structures from experimental data about human interactions,  arguing ``that a set of more than two people form a clique if they are all mutual friends of one another''.~\autocite[p. 97 f.]{luceMethodMatrixAnalysis1949} Although their matrix-based approach was less prone to errors than a cumbersome manual investigation of the data, the mathematical definition of a clique is often too restrictive in applications. Hence, later concepts in the different lineages of `community' can often be understood as weaker or looser versions of cliques that allow for sparser relations within groups.

In a review of community detection algorithms, Fortunato traces the origins of community detection back to a 1955 paper in sociometry by Robert Weiss and Eugene Jacobson, who proposed a method to deduce working groups from a matrix of work relationships in a complex government agency.~\autocite{fortunatoCommunityDetectionGraphs2010,weissMethodAnalysisStructure1955} Their method of finding groups by reorganizing the matrix representation of a graph (see Section~\ref{sec:Louvain} for a definition of the `adjacency matrix' of a graph) corresponding to a sociogram was first introduced by Elaine Forsyth and Leo Katz in 1946 who in turn developed the famous sociometric approach to groups introduced by Jacob Moreno in the 1930s.~\autocite{forsythMatrixApproachAnalysis1946,morenoWhoShallSurvive1934}

We can also trace origins of community detection in psychology and anthropology. In a 1956 paper in psychology, Dorwin Cartwright and Frank Harary used graph theory to introduce the concept of structural balance to describe ``configurations of many different sorts, such as communication networks, power systems, sociometric structures, systems of orientations, or perhaps neural networks''.~\autocite{cartwrightStructuralBalanceGeneralization1956} The image of the later broad applicability of the techniques concerned can be glimpsed here.  Harary, who was a mathematician at the University of Michigan, was interested in the translation of social science concepts into graph theory and later also worked on applications in anthropology, where he developed clustering methods for signed graphs to study homophily.~\autocite{hageStructuralModelsAnthropology1984}

Yet another thread of the lineage is formed by the use of what are called `stochastic block models' that find their origins in the social science literature from the 1970s.  For a review of this very wide field see an overview by Lee and Wilkinson.~\autocite{leeReviewStochasticBlock2019} In general, stochastic block models provide notions of `structural equivalence' in graphs where the `role' of a node is determined by its link structure. Deterministic models were first introduced by a group of sociologists around Ronald Breiger in 1975~\autocite{breigerAlgorithmClusteringRelational1975} and stochastic models by Paul Holland et al. in 1983.~\autocite{hollandStochasticBlockmodelsFirst1983}

A common feature of the techniques developed in the social sciences described above is their shared goal of determining structurally similar nodes in graphs to identify individuals in social networks playing similar roles. %
However, we want to emphasise that social scientists from the different lineages described above did not use the term `community'. Other terms like `cohesive subgroups'~\autocite{wassermanSocialNetworkAnalysis1994} or `balance and clustering phenomena'~\autocite{hageStructuralModelsAnthropology1984} were used instead, each meaning different things. Moreover, a limiting factor for the development of community detection algorithms in the social sciences was the absence of  computational power in the early years of social network analysis, where algorithms had to be performed manually in a cumbersome process.

As social network forms become significant in how people understand society, Mayer argues that they effectively become ``behavioural instructions''~\autocite[p. 54]{mayerSociometrySearchEngines2009}. It is these ``instructions''—before the advent of their machining in social media—that also provide the grounds for another current of work that sets out approaches in which the idea of the network or a set of contacts has become something that is more self-consciously to be used or manipulated in order to achieve certain political ends or social benefits.  Work such as Manfred Kochen and Ithiel de Sola Pool's ``Contacts and Influences'', a manuscript circulating from the early 1950s and published in 1978~\autocite{desolapoolContactsInfluence1978}, Stanley Milgram's 1967 direct experimental work~\autocite{milgramSmallWorldProblem1967}, and Mark Granovetter's 1973 article ``The Strength of Weak Ties''~\autocite{granovetterStrengthWeakTies1973} exemplify this tendency.

The notion of ``weak ties'' addressed by such researchers was embraced in mathematical terms by Watts and Strogatz in 1998~\autocite{wattsCollectiveDynamicsSmallworld1998}.  One of the interesting aspects of such work that is the idiomatic kind of movement from the very specific to the general that it stages.  This work is predicated on a particular kind of social connection, a friendship, knowledge of or acquaintance with an other, a social link, the passing of information from one entity to another, as the key, indeed sole, unit of analysis.  It is predicated on a wager that from this base unit, if precisely logged, something larger can be agglomerated.  Whereas other approaches to understanding the social in mathematical terms have often worked on the basis of surveying or assembling a population as a statistics-yielding mass, to be probed by averages and the deviations that yield them, this work starts `from the bottom up' in a certain way by narrowly fixating on the choreography of  what each different method takes to be a link. It is this movement from the specific to the general that its enduring attraction also lies, and, it wagers, something like a community can be measured.

As far as we have been able to trace, the physicists Michelle Girvan and Mark Newman were first to use the term `community' to describe a computational object in network science. In a highly influential paper from 2002,  Girvan and Newman, who were both working at the Santa Fe Institute in New Mexico at that time, coined the term `community' in this context and also present what one might call the `founding articulation' of community detection:
\begin{quote}
    ``Consider for a moment the case of social networks---networks of friendships or other acquaintances between individuals. \textit{It is a matter of common experience that such networks seem to have communities in them}: subsets of vertices within which vertex-vertex connections are dense, but between which connections are less dense. [...] Communities in a social network might represent real social groupings, perhaps by interest or background''.~\autocite[p. 7821, our emphasis]{girvanCommunityStructureSocial2002}
\end{quote}
In this description of communities, Girvan and Newman call to the experience of other network scientists who have noticed similar patterns of dense subgraphs in social interaction networks before, to suggest that a metaphorical or ``commonsense'' framing of community can be translated into network science.\footnote{The term `community' was also coined as an alternative to `cluster', a popular notion to describe groups of points in computer science, because the `clustering coefficient' was already an established concept with a different meaning in network science.} While `community' refers to the groups of nodes, the problem of finding communities in networks is called `community detection'.\autocite{newmanNetworks2018} Interestingly, both terms were first introduced by physicists and not social scientists, but have become hegemonic since then.\footnote{The 2002 article by Girvan and Newman has become very influential in the field with 13,876 citations [as of May 2023] according to Semantic Scholar.~\autocite{ammarConstructionLiteratureGraph2018} Moreover, Santo Fortunato further popularised the term `community' with several reviews on the subject that have `community' in their titles,~\autocite{lancichinettiCommunityDetectionAlgorithms2009,fortunatoCommunityDetectionGraphs2010,fortunatoCommunityStructureGraphs2012} where he also justifies the concept of community by referring to social networks as ``paradigmatic examples of graphs with communities''.~\autocite{fortunatoCommunityDetectionGraphs2010}} 

With the increase in available computational power, researchers from statistical physics moved into the field of network science and started to use their own techniques, in particular the statistical modelling of real-world networks, the description of networks as historically evolving structures and a dynamical systems approach for studying dynamic interactions between nodes, culminating in a `new science of networks'.~\autocite{newmanStructureDynamicsNetworks2006} This also lead to some amount of re-invention of parts of the social science literature and the relabelling of concepts like `community detection' can thus be understood as a revival of social network analysis as `network science' under the influence of physics.\footnote{The popularity of Mark Granovetter's ``weak-ties argument''---the hypothesis that large-scale social network cohesion depends on weak links between different communities---among physicists has also contributed to their increased interest in network science and physicists referencing Granovetter ``might be considered
innovators in [the complex network] community''.~\autocite[p. 21 ]{keucheniusAdoptionAdaptationComputational2021} To test Granovetter's hypothesis, physicists analysed community structures of real-world networks and, for example, Jukka-Pekka (JP) Onnela et al. showed that in mobile communication networks ``removal of the weak links will delete the bridges that connect different communities''.~\autocite[p. 7334]{onnelaStructureTieStrengths2007}} Only afterwards did some of the discourse in physics link up with the social science contributions, although Newman wrote several papers with social sciences references to address this less cited literature.\autocite{freemanGoingWrongWay2008}

\begin{figure}[ht]
    \centering
    \includegraphics[width=\textwidth]{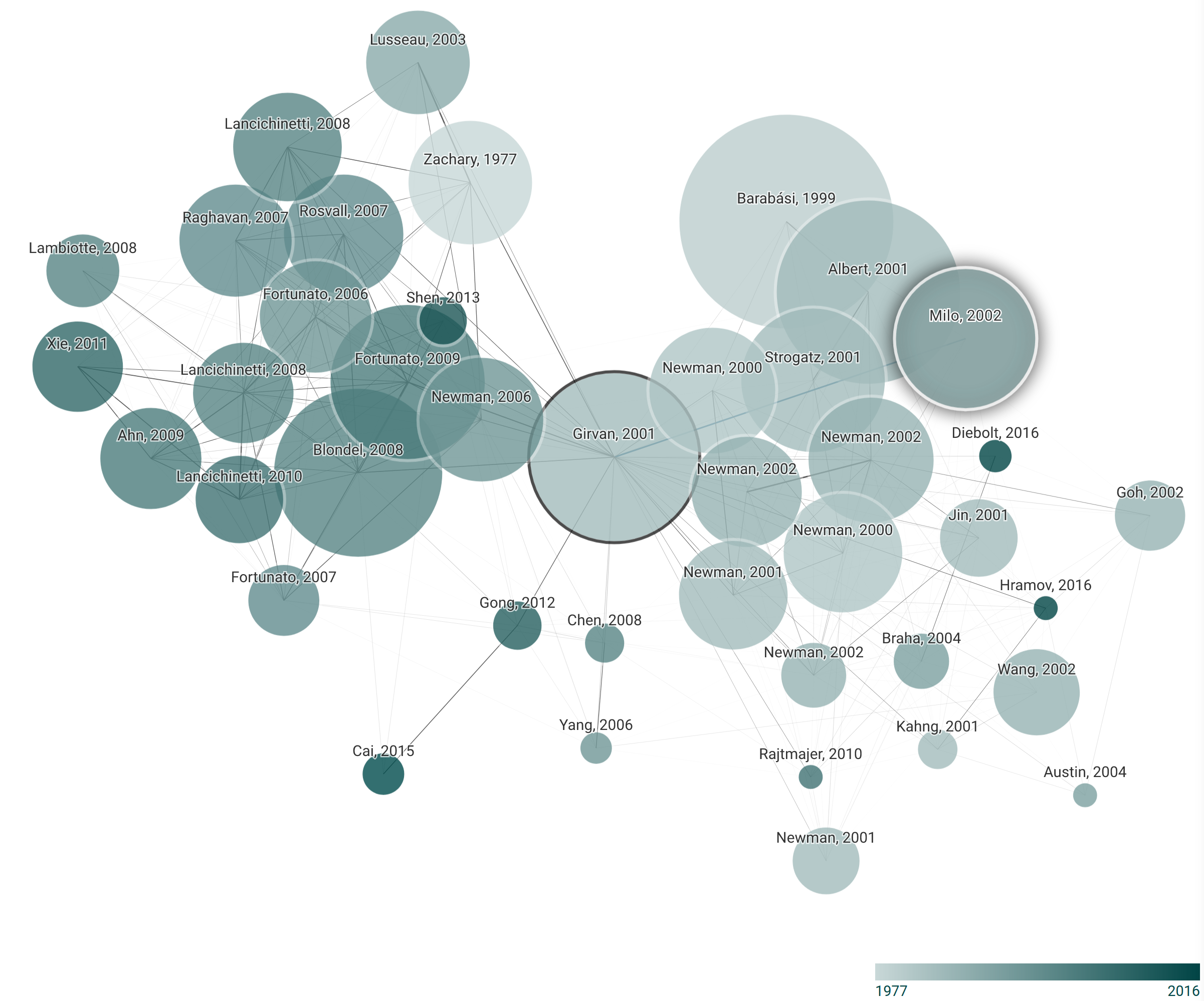}\caption{\textbf{Bibliographic graph for Girvan and Newman's original paper produced with the `Connected Papers' tool.} A bibliographic similarity graph is obtained for the original paper by Girvan and Newman placed in the center, using the online tool `Connected Papers' in April 2022. Nodes in the network represent the most important prior or derivative publications and edges are drawn according to similarity between papers, as determined by the `Connected Papers'. We observe that the influence of the social science literature on the field is not reflected in this bibliographic graph.} 
    \label{fig:citation_graph}
\end{figure}

After the formulation of the community detection problem, many functions were developed for computationally evaluating the quality of network partitions. One such evaluation function is that of `modularity' proposed by Newman and Girvan, who developed a framework of `modularity optimisation' to quantify the density of edges within communities as compared to a random edge configuration model.~\autocite{newmanFindingEvaluatingCommunity2004,newmanModularityCommunityStructure2006} Since then, a large body of literature has evolved and there are substantial investments in the maintenance and development of techniques for community detection.~\autocite{fortunatoCommunityDetectionGraphs2010} However, the field remains contested and recent approaches, such as `inferential methods' drawn from Bayesian statistics, challenge traditional techniques like modularity optimisation, which some authors even depreciate as merely `descriptive'.~\autocite{peixotoDescriptiveVsInferential2022,jacomyTwitterControversyCommunity2021} Despite such controversies, the modularity optimisation based `Louvain algorithm', introduced in 2008~\autocite{blondelFastUnfoldingCommunities2008} and named for the university at which it was developed, remains one of the standard methods for community detection. We provide a short analysis of the operation of this algorithm in the Section~\ref{sec:Louvain}.

There is an enormous bibliography associated with community detection which makes a detailed genealogy of the field difficult, especially because of the overlapping historical lineages we described above. Similarly, Fortunato suggests that ``the field has grown in a rather chaotic way, without a precise direction or guidelines''.~\autocite[p. 161]{fortunatoCommunityDetectionGraphs2010} A network science approach to this task could be to analyse the citation graph of, e.g., all scientific publications that cite the original article by Girvan and Newman and other works in the canon. A rigorous study of such citation graphs could be performed using the Academic Graph API by Semantic Scholar~\autocite{ammarConstructionLiteratureGraph2018} and applying community detection on this citation graph of community detection papers would be an amusing meta-exercise, although lying beyond the scope of this paper.\footnote{We notice that `friendship networks' of scientists are present in the stories of the genealogy of the field. Therefore, a network analysis of the collaboration and institutional affiliation networks of researchers working on community detection would also be illuminating to disentangle the different lineages from physics and the social sciences.} We restrict ourselves here to presenting a bibliographic graph obtained with the `Connected Papers' online tool based on the Academic Graph API that claims to construct a similarity graph consisting of the most important prior and derivative publications starting from a paper specified by the user.~\autocite{eitanConnectedPapersFind2022} Figure~\ref{fig:citation_graph} visualises this bibliographic graph as produced for the original paper by Girvan and Newman,~\autocite{girvanCommunityStructureSocial2002} which is placed at the centre. We observe that the graph over-represents the influence of physics literature on the field and renders the early influences from the social sciences that are outlined above invisible. Hence, the diagram constitutes an interesting cultural artefact that shows the current hegemony of physics in the network analysis field despite its partially sublimated but non-contiguous `origin' in the social sciences.

What has transformed over the course of the manifold set of lineages outlined in this section?  Aside from the changes in size of the networks to be graphed and of computational power, the transition has also entailed a general loosening from specifically addressing ideas of the social.  We can say that the techniques move from an abstraction from social groups involving techniques of observation, recording, encoding, and analysis into graphs that offer the production of a more generalisable formal structure.  This form then provides a means of addressing many different kind of entity.  It also provides a conceptual scaffold and technical substrate for new kinds of social relation to be grown.

\subsection{Notions of Community in Network Science}\label{sec:CommunityNetworkScience}

To understand how community figures in network science we continue by approaching the concept of community as used by network scientists whose mathematical definitions rely on graph theoretic formalisms. For a very comprehensive introduction to network science, the reader is referred to Katharina Zweig's book `Network analysis literacy',~\autocite{zweigNetworkAnalysisLiteracy2016} where the use of mathematical formalism is promoted because it abbreviates statements and makes them less ambiguous. %

\begin{figure}[ht]
    \centering
    \includegraphics[width=\textwidth]{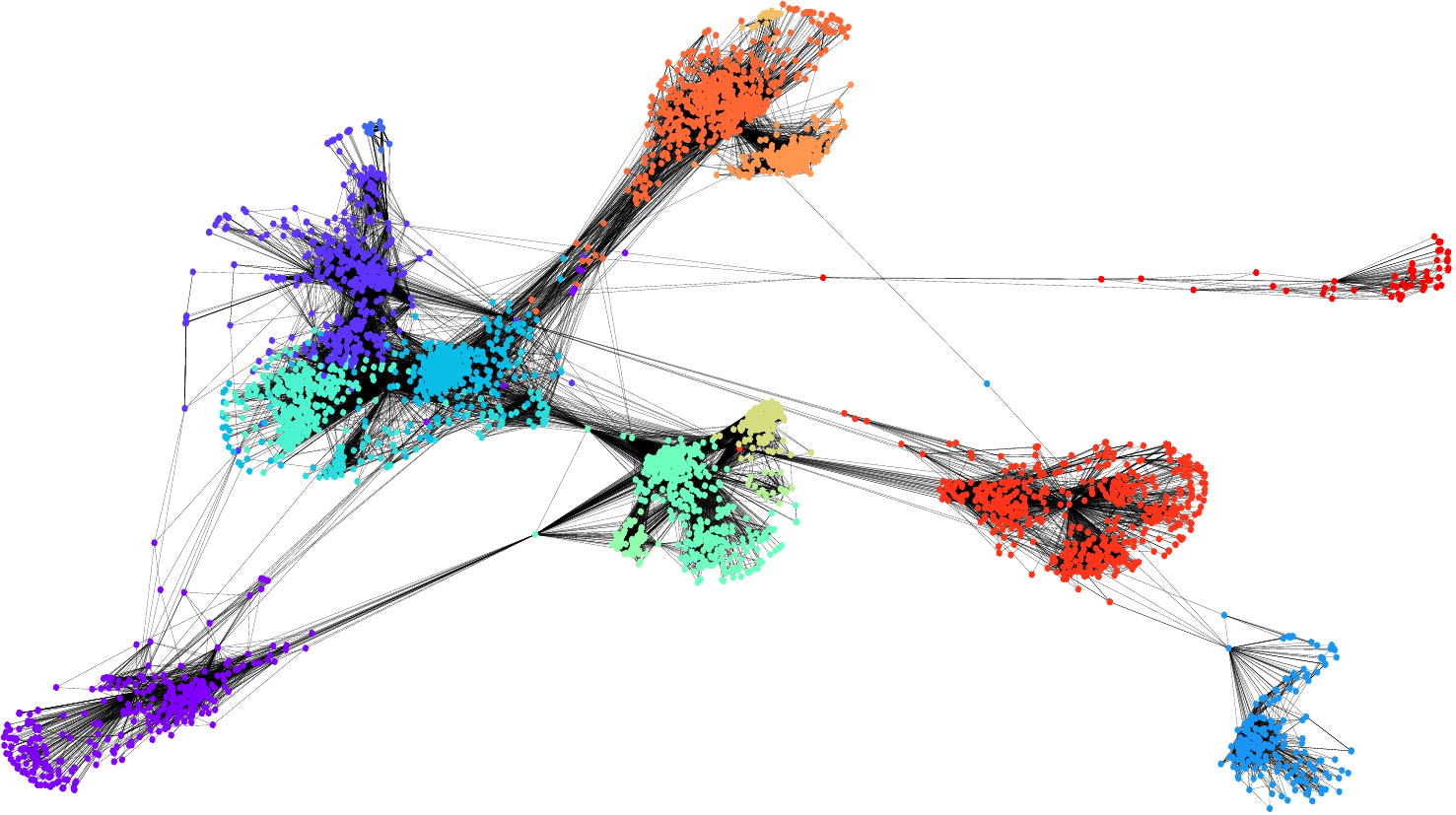}\caption{\textbf{Community detection in a Facebook friendship network.} We visualise the social network $G=(V,E)$ where vertices $i\in V$ (the points in the diagram) correspond to Facebook users and edges $\{i,j\}\in E$ (the lines in the diagram) are drawn between users that are `friends' on Facebook. Community detection using the Louvain algorithm is applied to the friendship network and nodes are coloured with respect to their assigned community, which network scientists would interpret as a friendship group. The network is derived from annonymised data provided by Julian McAuley and Jure Leskovec and we use the Python package `NetworkX' for computations and visualisation. We present this visualisation as an example of the diagrams produced in network science.}
    \label{fig:Example_Network}
\end{figure}

To give a flavour of the network science formalism, we guide the reader through some of the basic definitions with a running example at hand---a friendship network derived from anonymised Facebook data provided by Julian McAuley and Jure Leskovec.~\autocite{mcauleyDiscoveringSocialCircles2013} Note that detailed reading of the mathematical formalism in the following is not essential for the understanding of this article and we only come back to it in our case study in Section~\ref{sec:Louvain}. We use the Python package `NetworkX' for computations and visualisation,~\autocite{hagbergExploringNetworkStructure2008} and Figure~\ref{fig:Example_Network} depicts the example network that we denote by the symbol $G$. The network consists of $4,039$ Facebook users that are represented by points in the diagram called vertices (or nodes) and of $88,234$ friendships between users represented by lines in the diagram called edges (or links). To formalise this mathematically, each vertex is assigned a unique integer from $1$ to $4,039$ and the collection of all these distinct integers composes the \textit{set of vertices} that we denote by the symbol $V$. For an integer $i$, e.g. $i=1$, the notation $i\in V$ signifies that $i$ is a vertex in the graph $G$ or equivalently, an element of the set of vertices $V$. Accordingly, the symbol $\in$ is read as `element of'. To encode the relations between the vertices, i.e. the social relation of friendships in our example, we define an (undirected) edge between vertices $i$ and $j$ denoted by the symbol $\{i,j\}$ whenever $i$ and $j$ are `friends' on Facebook. The \textit{set of edges}, i.e. all friendships in our example, is denoted by the symbol $E$. Hence, our network $G$ consists of a finite set of vertices $V$ and a set of edges $E$ and this is often summarised by the notation $G=(V,E)$.\footnote{For the sake of simplicity we only defined the simplest form of networks here, the undirected and unweighted graph, meaning that the edges between nodes have no orientation but are symmetric and all edges have the same importance. However, it is also common to direct the edges and to weight them by a number that signifies the strength or importance of the connection, see~\autocite{zweigNetworkAnalysisLiteracy2016} for further information.}  

For network scientists, the term \emph{community} now serves as an abstraction often used to describe a group of vertices that have stronger relations among themselves than to the rest of the network,~\autocite{fortunatoCommunityDetectionGraphs2010} e.g. they share many friends or they play similar roles in the network. In particular, a community is determined entirely by its edges. The usual assumption is that networks can be divided into mutually exclusive or highly differentiated communities, where the number of communities is intrinsic to the network, and hence communities are understood to form the network's building blocks. The collection of these communities as building blocks is also called a network \textit{partition} and algorithms that are designed to find such partitions into communities are called \textit{community detection algorithms}. Figure \ref{fig:Example_Network} visualises community detection with the so called `Louvain algorithm'~\autocite{blondelFastUnfoldingCommunities2008} (see Section~\ref{sec:Louvain}) applied to our example network and the nodes in the figure are coloured with respect to their assigned community. The communities in the social network derived from Facebook could now be interpreted as different groups of friends and as we saw in the previous section, the notion of community is indeed motivated by the study of the social in a way that emphasises its interpretation as a concatenation of networks. One problem here is that it remains unclear if and to what extent community detection really operationalises the sociological concept of `community' and while Girvan and Newman are cautious to make this claim~\autocite{girvanCommunityStructureSocial2002}, other researchers use community detection algorithms exactly for that purpose.\footnote{See for an example of such use that is by no means unusual~\autocite{zhangCommunityIssuesDefinitions2012}.}

Today, community detection algorithms find applications in data from disciplines as diverse as biology, computer science, engineering, sociology or politics.~\autocite{fortunatoCommunityDetectionGraphs2010} Despite the high interest in community detection, there is---perhaps surprisingly---no generally accepted definition of what a community exactly constitutes in network science and so Michael Schaub et al. suggest that ``community detection should not be considered as a well-defined problem, but rather as an umbrella term with many facets''.~\autocite[p. 1]{schaubManyFacetsCommunity2017} They distinguish four different approaches:\footnote{We refer to Schaub et al. for a close-reading of the underlying mathematical formalisation of the four different approaches to community detection.}
\begin{enumerate}[(i)]
    \item Minimisation of violation constraints: When referred to as partitioning, the goal of community detection is to cut a network into several parts with the least cost, e.g. by breaking the least number of edges.
    \item Maximisation of internal density: In computer science, community detection is often treated as a discrete version of the data clustering problem and the goal is to find communities that have a very high internal density of edges but are less strongly connected across different communities. Girvan and Newman follow this clustering perspective.\autocite{girvanCommunityStructureSocial2002}
    \item Identification of structurally equivalent vertices: In the social sciences, community detection is often used to identify individuals in social networks that play the same role and so the goal is to find communities that consists of structurally equivalent nodes.
    \item Dynamic model reduction: Instead of focusing on the static network structure one can also study diffusion dynamics of e.g. information flows or an epidemic on the network and from a model reduction perspective, community detection then aims to determine a reduced graph that shows the same response to the dynamics as the original graph.
\end{enumerate}
While this summary gives an overview of different trends, there are also other categorisations of community detection, e.g. Santo Fortunato distinguishes `local definitions' of communities from `global definitions' and `definitions based on vertex similarity'~\autocite{fortunatoCommunityDetectionGraphs2010} or Tiago Peixoto divides community detection algorithms into `descriptive methods' and `inferential methods'.~\autocite{peixotoDescriptiveVsInferential2022} The persistence of multiple categorisations further underlines the way in which community remains an ambivalent concept in network science and cannot be reduced to one precise meaning. In particular, the translation of community into network science is not seamless and Newman himself  states that the description of community detection is ``vague and open to interpretation''.\autocite[p. 498]{newmanNetworks2018} Rather than seeing this as a problem however, we see that there are apt reasons for this relatively vague definitional state of the term.  Before elaborating this argument more fully, we map how it plays out in a key algorithm.

\subsection{Case Study: Louvain Algorithm}\label{sec:Louvain}

One popular technique in community detection methods is the so-called `Louvain algorithm' developed by Vincent Blondel and colleagues in the University of Louvain,~\autocite{blondelFastUnfoldingCommunities2008} which we will study in more detail here due to its high status and widespread usage. At the point of its development in 2008, the Louvain algorithm outperformed other popular methods for community detection with fast computational times and unprecedented scalability to extensive networks with more than 100 million nodes.~\autocite{blondelFastUnfoldingCommunities2008} Since then it has been widely used, e.g., community detection in the popular Gephi software~\autocite{bastianGephiOpenSource2009} for network analysis is implemented with the Louvain algorithm.~\autocite{gephiGephiTutorialQuick2010} Hence, many researchers from the computational social sciences---among whom Gephi is popular due to its accessibility~\autocite{jacomyGephiPaperGets2019}---ubiquitously rely on the Louvain algorithm for their study of social networks. 

While the Louvain algorithm remains very popular and has been picked up widely with more than 14,831 citations [as of May 2022] according to Semantic Scholar,~\autocite{ammarConstructionLiteratureGraph2018} there is a plethora of other community detection methods, made available by software projects like the python `Community Discovery Library' (CDlib) that has currently implemented about 100 different methods including Louvain.~\autocite{rossettiCDLIBPythonLibrary2019} Community detection in general, and the Louvain algorithm specifically, are thus moving targets for analysis and there are other algorithms in the field, e.g., the recently developed Leiden algorithm has some crucial advantages over the Louvain algorithm in terms of how it evaluates the connection quality within communities.~\autocite{traagLouvainLeidenGuaranteeing2019} The Leiden algorithm has thus gained in popularity among practitioners with Gephi planning to add it to their software.~\autocite{jacomyGephiCodeSustainability2021}  Nevertheless, since Louvain is so widely used and so extensively implemented it makes an ideal case study for understanding community detection.

The Louvain algorithm was one of the first algorithms to give an efficient optimisation heuristic for the `modularity' measure developed by Newman and Girvan.~\autocite{newmanFindingEvaluatingCommunity2004,newmanModularityCommunityStructure2006} With the help of modularity, usually denoted by the symbol $Q$, it is possible to compare the quality of different partitions and high modularity signifies good quality. Hence, the goal of a community detection algorithm based on modularity optimisation is to find a partition that maximises achievable modularity. We will introduce the mathematical formalisation of modularity and discuss some of its philosophical assumptions below. For the moment we remark that modularity optimisation follows the clustering perspective~(ii) on community detection as introduced in Section~\ref{sec:CommunityNetworkScience}. The Louvain algorithm is designed as a so called `greedy algorithm', which tries to find an optimal solution for modularity optimisation at each of its iterative steps.~\autocite{blondelFastUnfoldingCommunities2008}
We present pseudo-code of the algorithm below and describe its sequence of steps in the following.\footnote{Python code for the Louvain algorithm is available as part of the NetworkX package, see \url{https://networkx.org/documentation/stable/_modules/networkx/algorithms/community/louvain.html}} 
Note that both modularity and the Louvain algorithm can be applied to more general networks, in particular weighted and directed ones, but for the sake of simplicity we only consider unweighted and undirected networks here, as introduced in Section~\ref{sec:CommunityNetworkScience}. The input of the Louvain algorithm is a graph $G=(V,E)$. After initially assigning each node to a different community (step 1), the Louvain algorithm consists of two routines. In the first routine, one randomly loops over the nodes and a node $i\in V$ is added to a neighbouring community whenever the modularity $Q$ increases (step 4). After no further increase of modularity is possible with this strategy, a new `meta-network' is generated in the second routine, where the communities are defined as `meta-nodes' and the edges are aggregated (step 10).\footnote{The aggregation of edges in the meta-network is a non-trivial step that we cannot present in full detail here. In principal, edges within communities are aggregated to self-loops of the meta-nodes, where a self-loop refers to an edge starting and ending at the same node, and edges between communities are aggregated to edges between meta-nodes. However, the aggregation requires a weighting of the edges in order to represent the different edge densities correctly. We refer the reader to the original article for more detail.~\autocite{blondelFastUnfoldingCommunities2008}} 
After this aggregation, the first routine is again applied to the meta-network (step 11) and the meta-nodes are grouped together in communities by the two routines as before. These two routines are now iteratively repeated in a feedback-loop between input and output until no further increase in modularity is possible and the algorithm terminates (step 8). The resulting partition can be derived from the communities of meta-nodes and corresponds to a local maximum of modularity. The optimal partition determined in this way is the output of the Louvain algorithm. To phrase this in a different way, we set the algorithm out in pseudo-code below.

\begin{algorithm}
\caption{Louvain algorithm}\label{alg:Louvain}
\begin{algorithmic}[1]
\Require{Graph $G=(V,E)$}
\Ensure{Partition of $V$ into mutually exclusive communities}
\State Assign each node $i\in V$ to its own community.
\While{further improvement of modularity $Q$ is possible}
\For{each node $i \in V$ in random order}
\State Add $i$ to neighbouring community when modularity $Q$ increases.
\EndFor
\EndWhile
\If{all nodes are assigned to their own community}
\State Output the partition and \textbf{terminate}.
\Else
\State Construct meta-network $G_{\text{new}}$ with communities as meta-nodes and aggregate edges. 
\State Set $G=G_{\text{new}}$ and \textbf{return to step 1}.
\EndIf
\end{algorithmic}
\end{algorithm}

Figure~\ref{fig:Example_Network} visualises communities obtained from the application of the Louvain algorithm to our example friendship network, where nodes of the same colour are part of the same community. We used the Fruchterman-Reingold algorithm~\autocite{fruchtermanGraphDrawingForcedirected1991} to draw our network in the two-dimensional plane, which simulates forces of attraction between highly connected nodes and repulsion between disconnected nodes. Note that the close match between the force-directed network layout and community structure is not incidental. In fact, the Fruchterman-Reingold layout can be used to obtain communities via spatial proximity that optimise modularity~\autocite{songForceDirectedLayoutCommunity2013} and it was shown that force-directed layouts based on energy models can subsume the modularity measure.~\autocite{noackModularityClusteringForcedirected2009}

One could now obtain descriptive statistics to gain a better understanding of the community structure, e.g. the network is partitioned into 16 communities of which the largest (red) consists of 548 nodes and the smallest (light green) of 19.\footnote{We note that the colour gradient used in Figure~\ref{alg:Louvain} is unfortunately very fine due to the high number of communities. This can make the visual differentiation of communities challenging.} Note that one of modularity's features is that the number of communities is intrinsic to the network and can be recovered with the modularity optimisation. This is different to other data clustering algorithms like `$k$-means clustering', where the number of clusters has to be specified in advance \emph{a priori}.~\autocite{hastieElementsStatisticalLearning2009} We also remark that modularity optimisation is known to have certain technical limitations that have been well studied in the literature---most strikingly the `resolution limit' that prevents modularity finding relatively small community structures~.\autocite{fortunatoCommunityDetectionGraphs2010} Solutions to the different technical drawbacks of modularity optimisation have been proposed in the literature, e.g. the resolution-limit-free `Constant Potts Model',~\autocite{traagNarrowScopeResolutionlimitfree2011} but a review of these techniques lies beyond the scope of this article. 

To deepen our understanding of the Louvain algorithm, we now give an introduction to the modularity measure at its core,~\autocite{newmanFindingEvaluatingCommunity2004,newmanModularityCommunityStructure2006} for which we require an additional mathematical formalism from graph theory.\footnote{The graph-theoretic formalism used here mainly draws from linear algebra and matrix theory because they allow for an easy description and precise manipulation of graph structures~\autocite{zweigNetworkAnalysisLiteracy2016}.} We first introduce the so-called `adjacency matrix' of a graph, which is a very useful tool to represent the graph structure. For a graph $G$ with $N$ different nodes, e.g. our example graph of friendships on Facebook from Section~\ref{sec:CommunityNetworkScience} has 4,039 nodes, we define the adjacency matrix denoted by the symbol $A$ as a quadratic matrix, i.e. a matrix with the same number $N$ of rows as of columns. The entry of the adjacency matrix at row $i$ and column $j$ is denoted by the symbol $A_{ij}$ and represents a binary encoding for the presence or absence of an edge between nodes $i$ and $j$.~\autocite{zweigNetworkAnalysisLiteracy2016} More specifically, we define $A_{ij}=1$ if there is an edge between $i$ and $j$ %
and $A_{ij}=0$ otherwise.
Next we define the degree $d_i$ of a node $i\in V$ as the number of edges attached to node $i$ and using the adjacency matrix the degree can be computed as:
\[\tikzmarknode{d}{d_i} = \sum^N
  _{\tikzmarknode{j}{j=1}}
   \tikzmarknode{A}{A_{ij}}, %
\begin{tikzpicture}[overlay,remember picture,teal,>=stealth,shorten
 <=0.2ex,nodes={font=\tiny,align=left,inner ysep=1pt},<-]

\path (j.south) ++ (0,-1.5em) node[anchor=north west] (u)
{\textbf{Summation:}\\ Sum over all nodes $j$ in the graph};
\draw (j.south) |- ([xshift=0.3ex]u.south east);

\path (d.north) ++ (-10em,1em) node[anchor=south west,xshift=1.2ex] (o)
{\textbf{Degree:}\\Number of edges attached to $i$};
\draw (d.north) |- ([xshift=0ex]o.south west);

\path (A.north) ++ (0,1.5em) node[anchor=south west,xshift=-1.2ex] (il)
{\textbf{Adjacency matrix:}\\ Binary encoding for presence of edge between nodes $i$ and $j$};
\draw (A.north) |- ([xshift=0.3ex]il.south east);

\path let \p1=($(o.north)-(u.south)$),\p2=($(j.east)$),\p3=($(d.west)$),\n1={\x2-\x3} in \pgfextra{\xdef\tmpvspace{\y1}\xdef\tmphspace{\n1}};
\end{tikzpicture}\vcenter{\vspace{\tmpvspace}}
\hspace{\tmphspace} 
\]

\ \\
\bigskip \\
\noindent
where the mathematical notation $\sum_{j=1}^N$ means that the sum over elements indexed by $j$ ranging from 1 to $N$ is computed. As each entry $A_{ij}$ of the adjacency matrix encodes the presence of an edge attached to $i$ and another node $j$, the sum in the above equation %
counts the number of edges attached to $i$ as desired. %
Let us further denote the number of edges in the graph by the symbol $M$, e.g. the example friendship network has 88,234 edges. Before we can finally give the formula for the modularity, we need to establish a notation that allows us to compare whether two nodes are part of the same community. On that account, we denote the community of node $i\in V$ by the symbol $C_i$ and if two nodes $i$ and $j$ are in the same community this implies $C_i=C_j$. %

For a given partition of the network into different communities, the modularity denoted by the symbol $Q$ then measures the density of edges within communities as compared to a random rewiring of edges. Following the presentation by Blondel et al.~\autocite{blondelFastUnfoldingCommunities2008} and using our notation developed above, we finally define the modularity $Q$ as:
\
\bigskip \\
\[\tikzmarknode{d}{Q} = \sum^N
  _{\tikzmarknode{j}{\substack{i,j\\C_i=C_j}}}
   \left[
   \tikzmarknode{A}{\frac{A_{ij}}{2M}}-\tikzmarknode{p}{\frac{d_i d_j}{4M^2}}\right], 
\begin{tikzpicture}[overlay,remember picture,teal,>=stealth,shorten
 <=0.2ex,nodes={font=\tiny,align=left,inner ysep=1pt},<-]

\path (j.south) ++ (0,-2em) node[anchor=north west] (u)
{\textbf{Summation:}\\ Sum over all pairs of nodes $i$ and $j$ that are in the same community};
\draw (j.south) |- ([xshift=0.3ex]u.south east);

\path (d.north) ++ (-12em,1em) node[anchor=south west,xshift=1.2ex] (o)
{\textbf{Modularity:} \\ Measure for the quality of a partition};
\draw (d.north) |- ([xshift=0ex]o.south west);

\path (A.north) ++ (0,1.5em) node[anchor=south west,xshift=-1.2ex] (il)
{\textbf{Graph edge density:}\\Ratio of edges between nodes $i$ and $j$ in the graph};
\draw (A.north) |- ([xshift=0.3ex]il.south east);

\path (p.south) ++ (0,-1.2em) node[anchor=north west] (u)
{\textbf{Randomised edge density:}\\ Probability of edge between nodes $i$ and $j$ after random rewiring};
\draw (p.south) |- ([xshift=0.3ex]u.south east);

\path let \p1=($(o.north)-(u.south)$),\p2=($(j.east)$),\p3=($(d.west)$),\n1={\x2-\x3} in \pgfextra{\xdef\tmpvspace{\y1}\xdef\tmphspace{\n1}};
\end{tikzpicture}\vcenter{\vspace{\tmpvspace}}
\hspace{\tmphspace} 
\]

\
\bigskip \\
\bigskip \\
\noindent
where the sum is executed over all pairs of nodes that are part of the same community. While the first term in brackets corresponds to the ratio of edges between nodes $i$ and $j$, the second term computes the probability of an edge being present between nodes $i$ and $j$ after a rewiring of the graph that only preserves the node degrees.

We summarise that in the modularity-optimisation approach as realised by the Louvain algorithm, a community constitutes a group of nodes with a high density of internal edges.  In particular, neighbouring nodes are said to share a community when their connection is stronger than it would be were links to be simply randomly rewired.  Membership in a community is assumed to be objective and inherently described by node attributes such that an inquiry of the underlying entities replaced by nodes in the network framework (e.g. of the people behind the Facebook accounts in our example friendship network) becomes unnecessary. Moreover, membership in a community is assumed to be a static and unique node attribute and this leads to an essentialisation of community membership as a node feature. Hence, community detection applied to social networks contrasts with a view of a community as something that emerges from imaginal and material social relations shaped by historical and ecological conditions. Community detection, for the purpose of finding communities in social networks, certainly entails the reduction of complex social relations into sets of discrete edges in a network. Still, community detection is widely appreciated as a `good enough' tool and this leads us to analyse community detection algorithms as a form of heuristics in Section~\ref{sec:Heuristics}.  In order to understand why this might be useful, we need to understand the vagueness of community as a term.

\section{Vague Operators}

\subsection{Community as a Vague Operator}\label{sec:VagueOperator}

Although the algorithms designed for community detection mostly build on an \emph{ad hoc} and non-committal understanding of a community, the term community also refers to a complex sociological concept that is often politically loaded and whose understanding can abruptly differ between, say, different scholarly or activist traditions, amongst other kinds of uses of the term. The word community thus serves as a conceptual boundary object that mediates, for instance, between the social and the computational sciences. The way that it does so is however quite idiosyncratically nebulous, in a way that invites discussion.

Key to Susan Leigh Star's and her collaborators' notion of the boundary object as it iterated over time and across different cases, is the idea of \textit{interpretative flexibility}, that the same object can be read in many different ways, or for different purposes.  An example would be a map of an area, which can be read to plan different journeys, or to carry out operations as diverse as strategising a military campaign, allocating goods or services, or retrospectively tracing the possible routes of a vehicle, amongst many others.  Key to such examples is the way in which the precision of the boundary object enables multiple non-exclusive acts of interpretation.  In some ways this makes it like other modern phenomena such as an ideal ``writerly text'',~\autocite{barthes1974} or what curator Lawrence Alloway called the ``multi-evocative''~\autocite{allowayEduardoPaolozzi2011} .  Star's formulation is aimed at discerning \textit{cooperation without consensus} where an object is shared or generated by users who don't necessarily agree on its nature or what it is for. By comparison, some of the tension between a formalism (like ``community'') and its uses can be found in another context, one quite similar to that of certain implementations of network science—in that users may, as their name implies, see a use in a network, whilst the data on their interactions are the subject of exchange—is the distinction between exchange value and use value. In this economic faceting, the political contestation embedded in the epistemic comes to the fore.  The exchange of a thing only partially determines or encodes its use or interpretation, and more fundamentally, the epistemic formations that go into the genesis of such an entity. Nevertheless, at times such factors can be strongly determining.

In an article reviewing the way in which the boundary object term has been disseminated—and it has been justly influential~\autocite{starThisNotBoundary2010}—Star notes aspects of the proposal that have been less widely taken up. One of these is for vague uses that as she says are ``NOT interdisciplinary'' that is, that whilst such objects might be mutually used or generated by different disciplines, they are not themselves the direct grounds for interaction between the disciplines concerned.~\autocite[pp. 604-5, capitalisation in original]{starThisNotBoundary2010}  Star also notes the vagueness of aspects of some boundary objects as a condition that enables their usefulness in certain contexts.~\autocite[p.607]{starThisNotBoundary2010} It is this aspect that is applicable here.  Community is a central yet vague concept in network science, and it is this vagueness that comes to have subtle importance. This can be nuanced in two ways.  Firstly, community detection algorithms are not in themselves vague because algorithms are, in their own terms, precise yet `community' is overloaded as a term, because different `community detection algorithms', each of them precisely defined and implemented, produce or detect different network communities. The differences between them may be subtle or radical depending on their kind and the nature of the data that they interact with. Secondly, these different algorithms refer back to the term `community' as a kind of boundary object,  which holds many different meanings, but it is its affordance of vagueness that becomes productive.

In an earlier article (on a museum of zoology) Star and James R. Griesemer develop the idea of the boundary object to rework some of Bruno Latour, Michel Callon and John Law's formulations around contests of meaning in the intersection points of `diverse social worlds', where there is competition over the meaning of terms and the establishment of the means by which interpretations and significance can be established.~\autocite{starInstitutionalEcologyTranslations1989}  The boundary object becomes more or less fixed but only through what they call ``ecological'' means which are dynamic and multi-dimensional tussles over terms and practices.  By contrast, in the uses of the term community that we encounter in network science, there is a rather different way of operating.  There is an indifference to recruiting other users as allies in the same usage of the word.  We also note that there is sparse construction of ``obligatory points of passage''~\autocite[p. 111]{lawTechnologyClosureHeterogeneous1987} in a mechanics formed by the interplay of `interests'~\autocite{callonInterestsTheirTransformation1982} since there is no competition on these grounds and no threat of displacement from one meaning to another without the risk of producing `interpretative flexibility' since the object is vague rather than precise.  Rather, than being fought over, `community' is used, in this context, as something like an aggregate of hints, none of which necessarily `cash out' as more than an atmospheric term or loose identifier of a broad category.  In this sense, community acts as a further kind of boundary object, that could be called, following Star's recognition of vagueness, a \textit{vague operator}.

Vagueness can be useful, that is, put to many uses, some of which are tangential to each other and are not answerable to each other. We aim to map some of these uses, noting the way that as a vague operator, the term `community' allows for different ideational, discursive, technical and mathematical operations to co-exist in an overlapping and often mutually indifferent way.  Such uses indeed indicate aspects of the nature of a community, but in the sense that they can also cut across and interfere with each other, create detours in meaning and derive different results, so we are also dealing with a community of meanings that may be linked by curious means.  As such, a vague operator is different to a boundary object because it may act as a means of inclusively obfuscating the terms of what is being conjoined, in this case, under the term `community'.  We do not mean to claim that this is done for nefarious means—although this is likely the case at times with a term as loaded as `community'—but more that by mobilising a term that has multiple simultaneous meanings, uncontested vagueness, rather than a clear boundary or a multidimensional struggle over meaning, may have certain kinds of effects. It is these that come to the fore in the literature under discussion.

At the same time as it is a vague concept, community is one that also has a conceptual allure since whilst it operates as a vague technical description in different scientific idioms, community also has many other dimensions to its meaning. The idea of community as a good, or as a unit of analysis, is also an object of desire in some sense, not as an object, but what is sought after---as a \textit{condition} of value. One kind of tension to bear in mind however would be that exemplified in the difference between the data on a social media platform being seen by the platform owners as units of analysis and exchange whilst being understood by the platform's users as elements of use.
    
Furthermore, the term `community' also mediates between the designers of algorithms and software users who might bring their own equally \emph{ad hoc} and vague understandings of a community. As such, the term community comes with slippage when the computational object is seamlessly replaced with a sociological category in software systems that implement community detection algorithms. The often invisible translation of concepts from the non-digital to the digital world can be problematic as Richard Harper et al. note: ``Boundary objects succeed when they allow both sides to get on with their concerns without interfering with the other. They start to fail when the clarity of this distinction blurs''.\autocite[p. 4]{harperWhatFile2013} Vague operators, by comparison to other kinds of boundary objects, perhaps start to succeed when the clarity of the distinction blurs, allowing different kinds of claim to be made without too many questions being asked about their nature. This is not to imply that they are to be treated as inherently loaded, or that their use implies a sleight of hand.  Rather, that their vagueness is the result of a certain kind of precision available to mathematical description that makes the words associated with it have a rather secondary quality. This factor indeed tilts the axis of our discussion away from that between boundary objects and vague operators to a broader consideration of the nature of mathematical knowledge in action.

In his essay on the Semiotics of Mathematics Brian Rotman points out that mathematics has the unusual property that ``its signs seem to be constructed [...] so as to sever their signifieds, what they are supposed to mean, from the real time and space within which their material signifiers occur''.~\autocite[p. 5]{rotmanSemioticsMathematics2000} This gives mathematics its unique and highly valuable ability to be both about anything at all—in that, one way or another, anything can be described, however partially,  by number—and, with equal intensity, mathematics is also able to perpetually rework itself through unlimited terrains of abstraction since the one thing a number can be about most acutely, is other numbers. In the case of community detection this means that the behaviours of systems typically studied by physics if described in certain limited and mathematical ways can also produce descriptions or mappings that are transferable to, or relevant in, other contexts, such as social ones. 

A further dimension here is that many of the things treated as communities by such approaches do not pre-exist in a non-digital state. They are  natively digital as Richard Rogers puts it to describe processes and entities that come into being in computational systems.~\autocite{rogersDigitalMethods2013} But, as such, they are also part of a wider category—the natively artificial. The artifice involved is not only computational, but also mathematical and cultural. Whereas calling something natively digital describes it in relation to a media, the natively artificial describes a state of genesis and possibility. Thus, the terms of community's artificiality deserve probing and the recognition that they can perhaps be redesigned. Artificiality can provide grounds for a work of variation in which the presumptions and determinants of many kinds of communities can be reworked rather than being treated as an identifiable given. 

Indeed, under the name community detection, what might often be sought after is not community in the social sense, but a more general class of processual things of which a social community can said to be an example---things which in combination produce a unity. Processual things as a broader class of self-regulating entities entail complexes of relation whose interaction of parts emergently develop a contingent whole where sets of relatively simple entities combine to produce a more complex state.   This new state may be describable as another kind of entity or idiom---a change from nodes to networks for instance.  Community detection aims at divining the structure of such a movement.  In turn, numerous kinds of entities, including social media, e-commerce and other platforms, attempt to determine the underlying patterns of such emergence and by doing so to harness and entrain them to certain kinds of programme (see Section~\ref{sec:Problems}). Creating the right harness for the capacity of self-regulation, of something more complex to come out of the interaction of simpler states of things as they emerge into other states, is, as Alexander Kluge and Oskar Negt argue in ``History and Obstinacy'' something that characterises much of the work of politics, psychology and economy. \autocite{klugeHistoryObstinacy2014} Their book charts the ways in which technologies address, grip and reframe human bodily, psychic and social capacities starting with such states of emergence. 

The quality of emergence, which is brought into community detection from those trained in physics as we saw in Section~\ref{sec:Genealogy}, is what is hankered for. Changes in the state of a thing, of apple juice into cider, a child into an adult, or the mutually coordinated arrangement of birds into a flock  are typical kinds of emergence.~\autocite{delandaMaterialistPhenomenologyPhilosophy2022} The mapping of an implicit collective into a graph marking simple transmissions of information and the ability of that mapping to register a movement towards a point at which these transmissions `go viral', betokens a capacity to abstract such emergence.  They are both moments of transition at which value of many different sorts can be gained. The craft of the algorithm writer~\autocite{sennettCraftsman2008}, or the user of a community detection algorithm, in this context is to find a way of cleaving to the moment of developing self-regulation or emergence in a way that is meaningful.  It should allow emergence to be described, without killing it off. In the context of a graph used in social media this delicacy would be required to avoid creating something alienating yet whilst---in an echo of another kind of capture of energies, labour and materials that is not without violence and domestication---successfully `milking' it for meaning and value.  One might imagine other formations in which a successful `hack' lies in re-reading or transcoding information that is available `for free' in one context, but re-contextualising it via other abstractions. The art is indeed in engendering reciprocal exchanges in which forms of feedback between the mapped entities and the forms of mapping themselves generate other conditions of emergence.
    
One of the things effected by this vagueness of the term community, but also its significance as a thing that may refer to value-bearing entities and processes is the distinction between the discrete and continuous typical of digital systems.\autocite{faziContingentComputationAbstraction2018} Here, vagueness facilitates the possibility of the idea that community is based on discrete connection rather than something describable in more continuous terms, or by means of translation, such as overhearing, innuendo or implication, a general sense of something.  At the same time, this movement can be reversed; nebulous continuities preceding the possibility of the discrete. The craft of the algorithm designer thus has something to do with recognising partiality, the inadequacy of description in a vague operator, as being potentially productive, whilst negotiating the way in which this quality also plays out in relation to other consistencies, such as available computational resource. To this end, an understanding of the development of algorithms involved in community detection such as we offer above is pertinent, as too is an engagement with the controversies running through them .

\subsection{Controversies arising from Community Detection}\label{sec:Problems}
It may seem as if we have earlier identified vague operators as some kind of idyllically woozy and indeterminate form that lends itself to a plurality of interpretation and use in a way that evades some of the problems of identification. However, what might pass as virtues in some contexts, or in another facet of the same broader context, can operate as difficulties in others, by providing different kinds of usefulness.  Controversies may arise in part due to the way the term community acts as a vague operator.

In this section we briefly identify some of the ambiguous applications of community detection, in particular `recommender systems' and `anomaly detection', and discuss problems arising with them. Here, the epistemic dimension of community detection as an approach to describing and implementing things in the world, creates particular interpretative tensions and imperatives as they interlace with certain kinds of contemporary power in their application. In such cases vague operators work in another way, as what Alfred Sohn Rethel~\autocite{sohn-rethelIntellectualManualLabour1978} and subsequently Alberto Toscano~\autocite{toscanoOpenSecretReal2008} name `real abstractions'. Real abstractions are a means of making a materialist account of ideas and formalisms as they enter into relations with other kinds of stuff, such as goods, persons, and economic structures. A crucial form of real abstraction is the instantiation of the split between exchange value and use value mentioned earlier where exchange value is the real abstraction.  It is called real since it has effects independently of the ideas we might have about it. Marx discusses the almost mystical state of the commodity or of money in its `pure' form, waiting to be exchanged, to be used.~\autocite{marxCapitalVolume1981}  We can say that mathematical entities, in the terms that Rotman sets up, have something of the same quality.  A number does not take with it any traces of its previous uses, it springs into the world each time afresh. This is what is often so pleasing about them, and to connect with another thread of this argument, makes them \textit{useful}, replete with affordances of possible use.  And they are also real abstractions in the sense that, as they migrate out of the ideational phase of mathematical practice and into relations with other parts of the world they start to gain traction on things, as explanations, indexes, veils and objects in themselves. Computer systems indeed can be seen as dynamic nested hierarchies of real abstractions, and as such they may embody and mediate tensions, potentialities and conflicts at other levels of abstraction.

\paragraph{Recommender Systems} 
Community detection is widely used in the design of recommender systems,~\autocite{gasparettiCommunityDetectionSocial2021} i.e. algorithms that recommend products in online marketplaces or prompt actions such as following a person on a social networking site based on a user's history on that site or, via third party means, across others. In particular, community detection (e.g. with the Louvain algorithm) is used to obtain groups of users with ``similar social characteristics''~\autocite{gasparettiCommunityDetectionSocial2021} to either directly perform link prediction,~\autocite{javedCommunityDetectionNetworks2018} where a user might be recommended to follow another user analysed as belonging to the same community, or to enable the application of more resource-intensive recommendation algorithms on smaller-scale communities.~\autocite{gasparettiCommunityDetectionSocial2021} Here, community detection tracks a history of relations between entities and connects to other mechanisms, such as cookies, that describe a state of operations: whether an action such as a purchase has been carried out, how long a browser window was open onto certain data, whether there is a return to particular objects or sources such as a playlist of tracks and so on. The use of recommender systems poses two problems that have been widely recognised, that of reinforcement or channeling, and that of reduction.
\begin{figure}[ht]
    \centering
    \includegraphics[width=\textwidth]{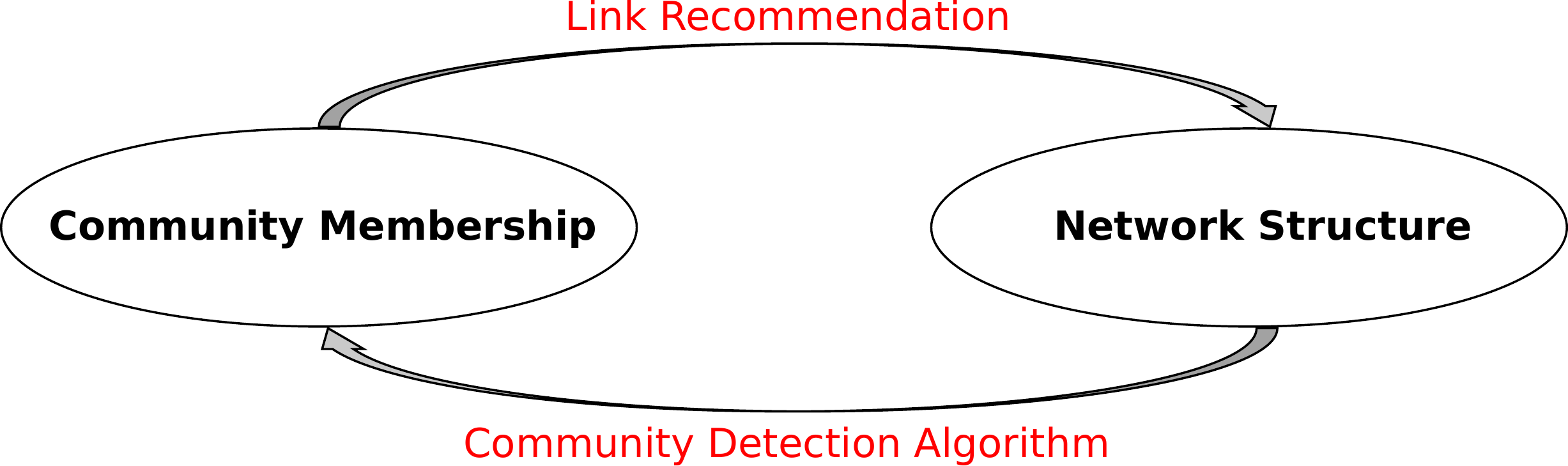}\caption{\textbf{Reinforcing loop of community detection and recommender systems.} Community membership leads to network structure by link recommendation based on recommender systems. From the network structure, new community memberships are deduced using community detection algorithms. Accelerated through this feedback loop, social networks become polarised and community detection might reinforce difference and segregation.} 
    \label{fig:reinforcing_loop}
\end{figure}
The first aspect of the problem acts as a kind of self-fulfilling prophecy (see Figure~\ref{fig:reinforcing_loop}), or what Dan McQuillan calls ``reinforcing loops'': ~\autocite{mcquillanResistingAIAntifascist2022} ever tightening loops between detection and recommendation that produce the by know well-known effects of filter bubbles and echo-chambers. But these effects also consist at another level in proposals to counter them by ``normative'' means that aim to rework them by a higher level of statistical generality. Bubbles and echo-chambers are anticipated and worked in advance, by approved parameters. Here  "corporate ethics" and regulatory policies, moderation systems, and other tricky apparatuses come into play.   The second aspect of this problem occurs through the operation of community as a form of real abstraction with jointly epistemic and material consequences.  Here, the biologist C.H. Waddington's notion of chreodisation or channelling is useful.~\autocite{waddingtonToolsThought1977} Waddington proposes attention to the ways in which patterns of development take place in a `landscape', which may be more or less recursively responsive, and which shape the range of actions that can be taken within it leading to the development of divergent and self-reinforcing branching structures or `chreods'. Herbert Simon's formulation of the evolution of ideas in a society~\autocite{simonRationalityTeleology1983} or organisation being partially determined by their niche and Sohn Rethel's Marxist arguments about the formation of epistemic entities within the dialectical formation of history~\autocite{sohn-rethelIntellectualManualLabour1978} are also relevant in mapping the kinds of processes in play. In all of these cases a reduction, for instance a sign or token, stands in for the real as a part of particular practices and activities.  Over time, or inherently to a process, that reduction is taken for the real and it is then used not only to refer to but to structure reality.  As a result, it may become or replace the real, depleting it.  This is of course a wider problem within modern societies where what might be called a ``reverse schizomogenesis'' occurs. In this sense, computing can tend in some ways towards a monoculture. To follow Yuk Hui's call for ``technodiversity'' we might try to counter such tendencies by recognising the diversities within mathematics and in mathematics as culture.~\autocite{huiQuestionConcerningTechnology2016}

Following Wendy Chun, who identifies `homophily', the principle that ``similarity breeds connection'',~\autocite{mcphersonBirdsFeatherHomophily2001} as one of the main drivers of `pattern discrimination' in network science,~\autocite{chunQueeryingHomophily2018} we can consider community detection as a method for performing `pattern discrimination'. Similar to homophily, the `performativity' of community is revealed in its capacity to draw boundaries and aggravate echo chambers in social networking sites as argued above. When following clustering perspectives~(ii) presented in Section~\ref{sec:CommunityNetworkScience}, one could interpret `community' as a reciprocal concept to `homophily' because 1) nodes in the same community are more densely connected and thus interpreted as similar and 2) similar nodes are expected to be densely connected and thus clustered in the same community. Community thus becomes a means for critically diagnosing homophily and also for implementing it. As a vague operator, it allows for entities in a network to exist in both of these complementary and mutually reinforcing states simultaneously whilst maintaining the duality of potential interpretations that also pulls them apart somewhat depending upon the interpretative niche or dialectical tension to which they are subject.  The vague operator of community here allows for the effective glossing of this state.

\paragraph{Anomaly detection}
An additional problem we want to note is one indicated by an explicitly political use of network science models of community. Matteo Pasquinelli proposes that algorithmic vision has two epistemic poles: of pattern recognition and anomaly detection.~\autocite{pasquinelliAnomalyDetectionMathematization2015} In network analysis, pattern recognition corresponds to community detection as described above. ``On the other side, anomalies are results that do not conform to a norm. The unexpected anomaly can be detected only against a pattern regularity''.~\autocite[p. 6]{pasquinelliAnomalyDetectionMathematization2015} Consequently, anomaly detection in social network analysis can also be performed with community-based network outlier detection methods that identify ambiguous nodes at the border between different communities.~\autocite{akogluGraphbasedAnomalyDetection2014} Moving between the poles of pattern recognition and anomaly detection, algorithmic vision within social network analysis aims to identify important actors in the network, their influence and interactions across different communities or to predict structural weak spots.~\autocite{scottSAGEHandbookSocial2011}
The network structure of digital control societies allows for extensive networked surveillance. Consider as an example the military Anomaly Detection at Multiple Scales (ADAMS) project funded by the US Defense Advanced Research Projects Agency (DARPA) with \$35 million that aims to surveille large-scale communication networks of e-mails and text messages for the identification of anomalies and security threats, especially whistleblowers:
\begin{quote}
``Each time we see an incident like a soldier in good mental health becoming homicidal or suicidal or an innocent insider becoming malicious we wonder why we didn’t see it coming. [...] ADAMS aims to rectify this situation by developing technology for the automated support of proactive use of the massive data sets being collected. [...] ADAMS will characterize graphs containing up to billions of nodes by structural feature sets calculated using recent breakthroughs in graph analytic techniques. ADAMS will use these features as the basis for novel anomaly detection algorithms.''~\autocite[p. 2f.]{darpaAnomalyDetectionMultiple2010}
\end{quote}
Here, the network graph is used as part of an apparatus aimed at establishing deep forms of control over behaviours and the graphed entities that exhibit them.  Different to the use of such techniques of identification in social media, the phenomena undergoing emergence here is one to be quashed rather than capitalised upon.  The reasons for defection from military logics are not examined, but optimised against.  There is a further lineage here, that of colonial and imperialist anti-insurgency techniques where populations are grouped and their connections are monitored.~\autocite{leporeIfThenHow2020} In both cases control over the identification and calculation of variation within what is established as a networked social form are key.  We can observe, in the passage cited above, that what might prompt variation from the enchannelled range of possible variation  is dubiously explained as a phenomena of mental health rather than of ethics or politics.  Good mental health, in behavioural terms such as these, becomes something identifiable through adherence to a set of norms of expression and behaviour.  Here, the notion of mental health operates as another kind of vague operator, but one identifiable through analysis of a `community' of actions, comunications and statements transcribed by the day-to-day working systems used by the militarised person.

These controversies exemplify some of the tensions in the computation of culture and society.  The push to forms of what are judged to be optimisation that they epitomise risks depleting the capacity for difference— in a form of devastation~\autocite{fullerBleakJoysAesthetics2019} via logical restructuring—that chains the much prized emergence that is sometimes limned by network analysis to a monotonous kind of ordering. The vague operator can play an enabling role in this. In order to mediate that condition, we want to go a little further to suggest a practical means of reframing the problem by reorienting a potential within it. How might it be possible to think and work critically with algorithmic processes whilst recognising the ways in which they are shaped by conditions of vague operation? One of the ways of doing so would be to think through the question of heuristics.

\subsection{Algorithms as Heuristics}\label{sec:Heuristics}

Blondel et al. call the Louvain algorithm a ``heuristic method that is based on modularity optimization''~\autocite[p. 1]{blondelFastUnfoldingCommunities2008} by following a `greedy strategy' of local optimisation at each step.  In the discussion of their paper they also suggest other `heuristics' (like thresholding for early stopping or excluding nodes with only one link from the analysis) to speed up their algorithm.~\autocite{blondelFastUnfoldingCommunities2008} Heuristics are used because the problem of community detection is `hard' and, in fact, it was shown by Brandes et al. that the task of modularity optimisation is indeed %
NP-hard.~\autocite{brandesModularityClustering2008} This means that there is no polynomial-time algorithm for exact modularity optimisation 
and this serves as a ``justification to use approximation algorithms and heuristics to cope with the problem''.~\autocite[p. 9]{brandesMaximizingModularityHard2006} In that sense, the Louvain algorithm is an `approximation algorithm' that gives an efficient heuristic for community detection via modularity optimisation. The more recent Leiden algorithm seeks to optimise the exact same quality function (modularity) but uses a different optimisation heuristic that produces better-connected components and runs faster.~\autocite{traagLouvainLeidenGuaranteeing2019}

These considerations of the Louvain and Leiden algorithms as heuristics serve as an impetus to study the deeper connections between algorithms and heuristics. We further propose that there is an affinity between operating via heuristics and understanding, configuring or playing along with something as a vague operator.

The term heuristics first arises in modern mathematics in the work of George Pólya who draws on ancient sources, and the early modern systematisers Descartes and Leibniz, to contextualise his study of mathematical problem solving. Pólya's 1945 book ``How to Solve It'' is a rich pragmatics of techniques for helping to get to solutions in a ``provisional and plausible'' way.~\autocite{polyaHowSolveIt2014} Pólya sees a heuristic as being of great use, but largely as a scaffold towards a rigorous proof. The term is taken up in the 1950s in the work of economist and artificial intelligence researcher Herbert Simon,~\autocite{simonModelsManSocial1957} where its potential application is broadened.~\footnote{In his memoir, Simon notes that his long-term collaborator, Allan Newell, who was a student of Pólya's introduced him to the term.  See,~\autocite{simonModelsMyLife1996}} In this incarnation it describes the kinds of economic decisions that can be made with limited information in the condition of what Simon called `bounded rationality'. A choice or decision always happens in some particular context, that of an organisation or administration, one of available data and the means to evaluate it, or a social context that drives and imposes certain notions of what is satisfactory. The question of what is satisfactory also prompts the development of another related term, `satisficing', where `satisfying' concerns finding a ``good enough move'',~\autocite[p. 205]{simonModelsManSocial1957} since, ``an organism that satisfices has no need of estimates of joint probability distributions, or of complete and consistent preference orderings of all possible alternatives of action.''~\autocite[p. 205]{simonModelsManSocial1957} 

The need for heuristics or techniques of satisficing in this area of research is set out in Simon's work on economic behaviour where he counters ``Olympian'' models of human reasoning~\autocite{simonReasonHumanAffairs1983} that aim to act from the possession of all facts. Heuristics are a way of working round this requirement for an absolute formal foundation, by making a more or less arbitrary `cut'.  The problem for a heuristic is to find an appropriate cut between smaller and larger amounts of information and information about that information. Approaches based in heuristics allow for approximate rather than absolute information to be the basis of a decision. It is important to note that the use of heuristics is not a question of being anti-reductivist.  Rather, the question is one of developing an effective technique that acknowledges the inevitability of reduction in the development of an abstraction or formalism and thus lowers the expectations of it from an ``Olympian'' scale, to a more pragmatic one. In a sense, heuristics could thus be understood as ``humble reductions''.  This formulation however returns the question of the scope and nature of the pragmatics involved and how it plays out in specific contexts and set-ups.\footnote{In military-funded work designed to address the construction of the automatic proof of logical statements Simon writes with Allen Newell to affirm that, ``all we are concerned with is that we have some criteria that `work' ''.~\autocite[p. 69]{newellLogicTheoryMachine1956} The pair further reinforce the position by arguing that, ``The method is a heuristic one, for it employs cues, based on the characteristics of the theorem to be proved, to limit the range of its search; it does not systematically enumerate all proofs. This use of cues represents a great saving in search, but carries the penalty that a proof may not in fact be found.  The test of a heuristic is empirical: does it work?''~\autocite[p. 71]{newellLogicTheoryMachine1956} A crucial part of the context within which satisficing must occur are the available quantities of computational resource and time, but also of the problem of trying to get something done to push forward a research agenda in a generally productive direction.} 

As Celia Lury has argued in a recent discussion of his work,~\autocite{luryProblemSpacesHow2021} Herbert Simon was particularly alert to the wider ways in which the construction of the form of efficacy has consequences for the formation of understanding. In common with many researchers working in frameworks developed after cybernetics, Simon experiments with linguistic and conceptual moves that seem to move between both naturalising technologies and technologising natural entities and processes, such as organisms; the concern is to tease out abstractions that work across these registers. Entities, whether ants, humans, organisations or societies are viewed as dynamic structures whose behaviour may be traced and modelled. These behaviours are seen as produced at the interface between the internal state of a system and its environment.~\autocite[Chapter 2]{simonSciencesArtificial1996} It may not be possible to fully describe all possible states of the interaction between such an entity and its environment, but approximation to them may be drawn up. Lury sets this work up as part of a disposition towards problems, drawing on the epistemological discussion of the problematic, but seen through the perspective of the particularities of the co-constitution of research methods and the material that they treat.

Algorithms of course don't work or even exist on their own. They are manifest in different modes of inscription or media, such as formulae or as pseudo-code for instance, and as something written into specific pieces of code running on particular systems at particular moments in time. They are further worked into and compose things such as social processes, database formats, and file structures amongst other things and in different contexts are something more or less indirectly experienced or undergone. They are worked through servers, by business plans and institutions, and are steeped in different kinds of politics of access. In turn, what is delivered by an algorithm producing a graph in a mathematical form requires further software to be visualised. Layout algorithms embedded in particular applications or libraries, are used—for instance, the Fruchterman-Reingold algorithm~\autocite{fruchtermanGraphDrawingForcedirected1991} mentioned in Section~\ref{sec:Louvain} —to place a diagram with intersecting edges on a two-dimensional plane.
The account of the Louvain algorithm offered in this article attempts to show such variation by articulating it in different forms, as too does the use here of different kinds of graphing embedded in specific libraries or particular informational products.  Each carries with it and sets-up different kind of imaginary and capacity to register or entail.

Here, Lury's formulations become particularly useful in that problematics are also intricately formed in techniques. The history of different rules of thumb or methods of approximation will themselves play out variably in different environments. Their particular acuities or fallibilities may become more marked, fraught, or less significant in different contexts. The relative size, granularity and qualities of data sets, the terms of the approximation, their fine detail and so on may provide significant aspects of such an environment, as might the more or less explicitly political dimensions of the milieus in which they form. Amidst all of this, the great usefulness of vague operators is partly in reworking the tensions between such scales and terms.  In this condition we need a critical heuristics to handle the multiply nested arrangement of such things in a landscape of vague operators.

\section{Gesticulations towards a critical heuristics of community detection}\label{Sec:CriticalHeuristics}

When a beetle topples over and lies upon its back waving its legs to the world, the arcs its limbs make in the air may have meaning, perhaps to the beetle, perhaps to entities in its environment, but those limbs may not meet resistance. Applied mathematics or network science, when its world is too easy, too readily arranged for its interpretation, may remind us of a beetle upon its back.  This resemblance is due to the conventionalisation of real abstractions, the too apparent applicability of many computing tools makes them verge on the edge of the illusory.  Computational ease tends to align with existing formations of power and concentrations of data-wealth that render certain things more possible than others, despite the great potential plasticity of computing.~\autocite{fullerPraisePlasticityUnderspecification2020} In a sense, we are back to Rotman's note on numbers as `severed signifiers' and the alternation between the grasp and construction of things by different forms of evaluation as variously decsribed by Lury or Sohn-Rethel. At the same time, such signifiers are immensely potent when they are not ``de-severed'', but plugged into problems that can be addressed by their capacity for abstraction. Network science can both contribute to setting up the conditions by which a population can be treated as tokens in a game of behavioural psychological warfare in advertising or political machinations as illustrated in Section~\ref{sec:Problems}, but it can also work to test the spread of an epidemic.~\autocite{brockmannHiddenGeometryComplex2013} Both of these are immensely useful, in certain ways, but simple facility is insufficient to judge their wider validity.

As we move towards the end of this article we want to draw on heuristics as a means of mediation between a formalism or an abstraction and those things such as data, or various forms of real on which, by means of various layers of translation and mediation, it gains traction or resistance. A heuristic is supposed to be a pragmatic cut, the result of a process of parlaying between the possible and the probable it is a `reasonable adjustment' subject to more or less equally reasonable doubt.  In the way that Blondel et al. use it in the design of the Louvain algorithm,~\autocite{blondelFastUnfoldingCommunities2008} heuristics are used in a way that acknowledge their own limits as a limited and contextualised exercise. What if a heuristic could also become speculative or critical rather than primarily pragmatic in the sense that it `works'? Some of the existing literature in network science and related fields prompts us to talk about this.

In their text, ``Clustering: Science or Art?''~\autocite{luxburgClusteringScienceArt2012} Ulrike von Luxburg, Robert C. Williamson and Isabelle Guyon suggest that clustering should be evaluated according to the downstream task where there are clearer-cut criteria for suitability. Similarly, in their article, ``Community Structure in Graphs'',~\autocite{fortunatoCommunityStructureGraphs2012} Santo Fortunato and Claudio Castellano argue that there is no `silver bullet' in community detection and no perfect algorithm exists for the task. This is formalised by Leto Peel, Daniel Larremore and Aaron Clauset who prove a `free lunch theorem' stating that community detection algorithms perform equally well when averaged over all possible problems and only on a subset of problems can one algorithm be preferred.~\autocite{peelGroundTruthMetadata2017} These different formulations all draw similar conclusions, if you want to do more than exploratory data analysis you have to tailor to the specific system at hand, or attend to the problem being addressed by producing more specialised algorithms.  The criteria for this evaluation might come from the problem being examined, but also from what counts in understanding the gestation of the problem and what counts as adequate means to formulate some kind of its knowing.

One mode would be to attempt to formulate a provisional ``quasi-universal'' in which the problem can be subsumed, to aggregate more information to dissolve the particulars of the problem in a conceptual substance supple and granular enough to absorb and rework its specificities in an enquiry driven by these. Another, is to separate out the problem, and only treat the immediate point in the established pipeline at which one is positioned through mental compartmentalisation and technical segmentation.  Another is to both critically engage with the epistemic and political dimensions of the problem at the same time as humbly and playfully working with heuristics. It is the latter option that we aim to gesticulate toward in four ways to imagine a `critical heuristics' that embraces partiality, works by means of epistemic humbleness, and offers capacities of reflexivity and artificiality. In this of course we recognise that heuristics, just like absolutely rigorous proof, on its own does not `solve' the problems within which mathematics is embedded.  At the same time, it may provide a means of examining the difficulties in, and reworking and moving amidst and across,  vague operations. 

\paragraph{Partiality}
It is worth teasing out some possible use of the heuristic mode. One way is to recognise the terms network and community themselves as forms of heuristic descriptor or rough approximation, both as heuristics and as vague operators with the different valences of interpretation that these terms imply. Often, in application, the term community can only be partially relevant, but as a vague operator it is more or less adequate,  and it is that partiality that is interesting. It is sometimes a partiality that does the work of convincing users that in some contexts it is a `good enough' description of a community as such, or at other times, the partiality that does the work of articulating sets of certain kinds of relations in something that is not quite a community, but that can be more or less usefully described as such.

What is  interesting here is the generative deployment of partiality: partiality as a form of productive discrepancy, partiality as constructive misreading, partiality as mistranslation layered upon others, partiality as a retraining of social forms into something `more computable'. In applied mathematics it is generally understood that such partialities are produced by idealisations or representations that to some extent always mistranslate. Partiality introduces the possibility for recursive operations of misplaced concreteness, as something that more or less maps onto something that has already been mediated into a graph.

Such partialities or approximations can make for a certain kind of forgiving or playful conviviality, but they have to be tuned into; as when Annie Ernaux talks of the ``approximate'' quality of conversations between lovers whose linguistic terrains only partially overlap.~\autocite[p.25]{ernauxSimplePassion2021} One can say this about such things as the different figures, formulae, analyses and modes of encoding that jostle together in this article.  Here, the romantic formulation of the illusory nature of exactitude (as for instance in Schopenhauer or Nietzsche) and its more recent reworkings comes into play. Partiality, being partial to something, can then be the grounds for congenial relations.  The patterns of interference between different modes and sample-rates of such partialities are also the grounds for a creative micro-politics, and, equally importantly, for the recognition of cruelties in what is rendered inexpressible, or what is sheared off.

\paragraph{Epistemic humbleness}
The vertex and the line act as forms of index, they refer to something such as a relation or an entity.  The idea of the index, even by something as severed as a number is a problem of representation. To establish this, we can ask, ``What would the movement of the beetle's legs look like in a frictionless medium?'' In other words, can we imagine an absolutely isomorphic mapping of an entity in a way that is only describable in the terms of edges and vertices, an object of enquiry that literally calls for such graphing because it can have no other manifestation? This is what exists in purely mathematical terms, but when they come into combination with entities and processes that they translate, some things may be missed, become vague, amplified or crossed out.  A critical heuristics would pay attention to what is rendered mute in such a condition, that which is lost, relations that are inexpressible or set to invisible, what is sloughed off as defiant of reason, rule or representability. It would recognise the crudeness of the abstractions it offers and work with them as a kind of Art Brut rather than as a revelation of Platonic verity or an unfortunate condition of ``things as they are''.  For instance, due to their ability to map relations only in one-to-one terms, one node to another in pairwise interactions,  it is difficult to express relations between individuals and collectives using network graphs.  Such a relationship might include that between a person and a state and its agents, such as the police, or between one class and another.  Graphs also cannot sense relations based upon exclusions, voids or devastations, what is not there.

A critical heuristics enquires into what entities warrant transubstantiation into nodes, which relations are describable as vertices, whose data is rendered accessible, what data is legible to which systems. It works as part of an enquiry into what is rendered economically interoperable through the perspectival operations of what is rendered as a perdurable glitch.~\autocite{russellGlitchFeminismManifesto2020} It looks incessantly for what is deemed to be outside the scope of the problem, that can provide resistance or footholds in terrains that may be social or political, ecological, conceptual or mathematical. (In this way concern with the ecological costs of computing might prompt a return to the proper veneration of terse and elegant algorithms and sparse use of computational resources for instance.) A critical heuristics might, in making propositions, subtly shift out of what can be the sometimes overly reactive trap of the critical mode. It would also shift the game of heuristics. One of the implicit claims of a heuristic is that it is a humble mode of thought and action.  Heuristics are humble because they acknowledge the reduction they make and recognise its provisionality and partiality.  Too often, this humbleness passes off as an excuse not to think as users bracket the provisionality of heuristics off when employing them. Might we imagine an \textit{actually humble} heuristics?

\paragraph{Reflexivity}
Instead of means to recognise the wider dimensions of relationality suggested in the note on humbleness we live amidst social graphs or office graphs that often use similar graphing structures not to bring more things into account in recognising the complexity of an event or a person, but to accumulate more kinds of things under the same representational regime and to entrain them to it. If we return to the use of heuristics in the case of the Louvain algorithm, whilst Louvain is described as a heuristic by its authors, its use might be quite different. The communities it graphs might be taken to be frictionlessly real, rather than something at least partially brought into being by the contact with and provision of resistance to the algorithm and the systems through which it operates. When communities are taken as simple reals it is tempting to forget the heuristic nature of their manifestation and the vague operation of their translation.

In order to interrogate this forgetting, often integral to the segmentation of a problem within modernity, there can be intuitively imagined a `pipeline' between different stages of the development of a technique. This pipeline, say, for an algorithm for network analysis, would run through: mathematics, where the approach is posed; computer science, where it is formalised; software engineering, where it is implemented; the domain of users, where it is deployed in software that they use to work on specific cases; finally, one end of the pipeline that occasionally appears is very much the last dribble of its flow, and is composed of critical readers or those who are seen as `complainers' who try to evaluate or `undermine' the entire effort, but whose work can be safely ignored at any stage of the pipeline.\footnote{This image of the pipeline should not be taken in an ethnographic sense, or one that endorses any hierarchy of knowledge or practices, but simply as an illustration of another kind of diagram, of socio-technical relations and practices. Actual conditions can involve more branching and looping, returning to a means to recognise the way a calculation comes into being through what it comes into composition with in the condition of emergence of a graph.} Our intent with this article is to suggest a closing of the loop of this piped construction and to work the epistemic and political analysis in with the mathematical, with software and its uses and interpretations. A critical heuristics might be helpful in making this line into a loop by introducing reflexivity into the pipeline such that practices are de-segmented and epistemically evaluated throughout, leading to a gurgling cascade rather than a streamlined flow.

Further, reflexivity can also be found in the form that, since social network analysis has become a `fact of life', through the various implementations of social media, or through dubious tools of control such as the London Metropolitan Police's ``Gangs Matrix'' or its successor systems,~\autocite{amnestyinternationalukTrappedMatrixSecrecy2018} and through citation analyses such as the h-index for academics, reflexivity is to be found simply in peoples' navigation of a social and informational domain in which these are operative and structurally significant factors.  There is scant opportunity for `naive' behaviour under such conditions, so network analysis can be said to map only those behaviours that are given under conditions of network analysis.  Reflexivity thus may also be found in the irony and cynicism induced by a metricised society of analysis.

\paragraph{Artificiality}
The problem of describing communities per se is not simply one of applying numbering techniques to something that naturally elides them. Numbers and numbering practices are not simply thin shadows of something already in the world that is more robust, meaningful and concrete, although they may sometimes  be so. Nor are they automatically reducible themselves to certain functions such as reification, this depends on the particular conjunction. Rather, they can also be recognised as something artificial and novel existing and working in the world irreducibly~\autocite{latourPasteurizationFrance1993}. Numbers have their own specific qualities, the uses and implications of which vary across, cultures, polities, technologies and implementation, in other words in historical terms.

Calculative practices and technologies can be inventive, arranging new entities and novel conjunctures. They also exist amidst thousands of other such things, allowing for the formation of computations traversing assemblages of different kinds. For instance, in this context the connection between modularity and community is one of these expressive conjunctions in that modularity is a function that is hard to optimise %
since it is NP-hard, whilst community is much vaguer. The overlap between these terms is slippy, a vague operation, the simultaneously ideational and material play in which offers itself to numerous kinds of use, a condition that demands epistemic and political work that embraces artificiality.~\autocite{simonSciencesArtificial1996} A heuristics that is able to play with and to speculate through this condition of artificiality is also one that is able to work through the pluralities of artificiality, including those that are deemed to be perverse, unwholesome, too recondite, obscure or geeky, that is, perhaps of decadent as well as inventive kinds.  It may also provide the grounds for testing the ways in which it is too readily put to work, to reclaim the productive fragility of precise knowledge in its dance with vagueness.

\section{Conclusion}

We propose that an expanded heuristics can provide a route to reflecting on the limits and wider dispositions of algorithmic knowledge. This motivates us to focus on community detection as a set of mathematical heuristics that can be used in ways that are potentially attuned to their limitations.  This combination of descriptive thinness and capacious applicability makes, we suggest, the term community a vague operator (a particular kind of boundary object).  Building on this, we propose a \textit{critical heuristics} in network science that has the capacity to both recognise and profit from its constructed nature and to proceed via humble epistemological claims. A heuristics that is more tactical, provisional and  contingent may also offer a recognition of the tensions and absences involved in such kinds of knowledge and real abstraction. In a sense we want to suggest that a `no free lunch' argument, in which every algorithm has its idiosyncratic costs and predelictions, can also be made at an epistemological and political level.  Here, the various kinds of interplay between the hyper-precise and the vague that are embodied in the conjuncture of particular algorithms and specific problems to be worked on are to be kept in mind as well as being implicitly mobilised in techniques.

\section*{Acknowledgements}
We would like to thank Michael Schaub and Mauricio Barahona for insightful discussions about community detection in network science and its different lineages. No errors here are ascribable to them. We would also like to thank the Centre for Digital Inquiry at Warwick University and the Digital Democracies Institute at Simon Fraser University for giving us the opportunity to present parts of this article in their respective seminars. Finally, we thank our anonymous reviewers for their very useful suggestions.
Funding in direct support of this work: DS acknowledges support from the EPSRC (PhD studentship through the Department of Mathematics at Imperial College London).

\setlength\bibitemsep{0pt}
\printbibliography

\end{document}